\documentclass[a4paper,pra,twocolumn,showpacs,superscriptaddress,groupedaddress]{revtex4}
\usepackage{ae}
\usepackage[T1]{fontenc}
\usepackage[ansinew]{inputenc}
\usepackage{amsmath}
\usepackage{amssymb}
\usepackage[caption=false]{subfig}
\usepackage{multirow}
\usepackage{array}
\usepackage[]{graphicx}
\usepackage{wrapfig}

\usepackage{color}
\usepackage[colorlinks]{hyperref}
\usepackage{lscape}
\begin{document}
\title{Performance of quantum cloning and deleting machines over coherence}
\author{Sumana Karmakar}
\email{sumanakarmakar88@gmail.com}
\affiliation{Department of Applied Mathematics, University of Calcutta, 92, A.P.C. Road, Kolkata-700009, India.}
\author{Ajoy Sen}
\email{ajoy.sn@gmail.com}
\affiliation{Department of Applied Mathematics, University of Calcutta, 92, A.P.C. Road, Kolkata-700009, India.}
\author{Debasis Sarkar}
\email{dsappmath@caluniv.ac.in, dsarkar1x@gmail.com}
\affiliation{Department of Applied Mathematics, University of Calcutta, 92, A.P.C. Road, Kolkata-700009, India.}
\begin{abstract}
Coherence, being at the heart of interference phenomena,
is found to be an useful resource in quantum information theory.
Here we want to understand quantum coherence under the combination of two
fundamentally dual processes, viz., cloning and deleting. We found the role
of quantum cloning and deletion machines with the consumption and generation
of quantum coherence. We establish cloning as a cohering process and deletion
as a decohering process. Fidelity of the process will be shown to have connection
 with coherence generation and consumption of the processes.
\end{abstract}
\maketitle

\section{Introduction}
No-cloning theorem plays a fundamental role in quantum information theory. The theorem states that an arbitrary quantum state cannot be cloned \cite{WZ} and this provides a fundamental support to the absolute security in quantum cryptography \cite{BH}. Although, unknown quantum states are forbidden in perfect duplication, some information about them can be obtained by suitable physical processes, such as unitary transformations. After cloning, the output copies should resemble to some extent to the copies of the states to be cloned and this resemblance is quantified by fidelity. There are two types of cloning machine. One is state-dependent cloning machine  and another is state-independent universal cloning machine. Wooters-Zurek cloning machine\cite{WZ} is state-dependent whereas optimal universal cloner\cite{optimal cloner} and phase-covariant quantum cloner\cite{Phase-cov cloner} are state independent. On the other hand, quantum deleting machine is applied in a situation when scarcity of memory in quantum computation occurs. No-deletion is also a fundamental no-go principle in quantum theory. It states that unlike classical theory, the perfect deletion of an unknown qubit from a collection of two or more qubits is an impossible operation. This was first observed by Pati and Braunstein \cite{deleting machine} where they showed that the linearity of quantum theory does not allow to delete a copy of an arbitrary quantum state perfectly in either a reversible or an irreversible manner. Deletion is a thermodynamically irreversible process. Quantum deletion is like `reversible copying' or `uncopying' of an unknown quantum state. Although `no-deleting principle' states the impossibility of constructing a perfect deleting machine \cite{deleting machine}, if quantum deletion could be done, then one would be able to create a standard blank state onto which an unknown quantum state can be copied approximately by deterministic cloning or by probabilistic cloning process \cite{cloning machine}. This deletion machine may be state dependent or state independent like cloning machine. Thus, cloning is a process by which we transfer information to a system and deletion is a process by which we can delete information from a system. Therefore, given cloning and deletion as two important dual processes, one would be interested to understand their role on different aspects of quantum information theory, viz., on resources like entanglement, coherence, etc.\\

Coherence, being at the heart of interference phenomena, plays a central role in physics as it enables applications that are impossible within classical domain or ray optics. The coherent superposition of states stands as one of the characteristic features that mark the departure of quantum mechanics from the classical realm, if not the most essential one. Recently, the theory of quantum coherence has attracted much more attention after some quantification schemes have been developed by Baumgratz et al. \cite{coh1}. Quantum coherence has well known for its essential role in biological systems\cite{bio1,bio2,bio3}, quantum thermodynamical phenomena \cite{thermo1,thermo2,thermo3,thermo4,thermo5}, quantum metrology, etc. It has been connected to entanglement and other quantumness measures like discord and deficit. In particular, both entanglement and coherence have their respective root in quantum superposition principle. There exist activation protocols which relate  generated entanglement with coherence. The resource theory of coherence has also been developed recently and it has been extended to multiparty level where entanglement has been shown to form a part of coherence. Different types of incoherent operations have also been identified and several new coherence quantification schemes \cite{coh1,coh2} have been introduced. 

Our main motivation in this work is to observe the effects of processes like, cloning, deletion on the resource like, coherence. As the processes, both cloning and deletion show their important connections with many aspects of quantum information theory e.g., there exists a strong analogy between quantum cloning and state estimation. Quantum cloning is equivalent to state estimation in the asymptotic regime where the number of clones tends to infinity\cite{state estimation1,state estimation2,state estimation3}. Quantum cloning also helps to detect any eavesdropper on a quantum channel, i.e., we could relate the security of quantum key distribution with cloning \cite{cryptography1,cryptography2,cryptography3}. Thus there are strong relationships between cloning and various important information processing tasks like state estimation, quantum cryptography, etc. On the other hand, quantum correlations are useful resources for these information processing tasks. These connections between cloning and as well deletion with various information processing tasks motivate us to analyze the behavior of quantum correlations under cloning and deleting operations. Such type of analysis for few non classical correlations like entanglement, discord have already been studied \cite{cloning-ent}. In this work, we will study the behavior of coherence, a new signature of quantumness, under the combination of two seemingly dual quantum processes - `cloning followed by deletion' and `deletion followed by cloning' and we will provide performances of both the processes
through quantum coherence consumption and generation and connect them with the fidelities of the processes. We  will establish cloning as a cohering process and deletion as a decohering process. Our paper is organized as follows. Section I contains introduction. In section II  and section III, we will discuss some background materials. Section IV and V contain our main results and section VI ended with the conclusion.

\section{quantum coherence}
Consider a finite dimensional Hilbert space $\mathcal{H}$ of dimension $d$. For a particular basis $\{|i\rangle\}_{i=1}^d$, let, $\mathcal{I}$ be the set of incoherent states. Mathematically, incoherent states are those states which are diagonal in the chosen basis. All incoherent density operators are of the form $\delta=\sum_{i=1}^d\delta_i|i\rangle\langle i|$ where $\delta_i\in[0,1]$ and $\sum_{i=1}^d\delta_i=1$. Baumgratz et al. \cite{coh1} have formulated a set of physical requirements which should be satisfied by any valid measure $C$ of quantum coherence.

\begin{enumerate}
  \item $C(\rho)\ge 0$ for all states $\rho$ and equality holds iff $\rho\in \mathcal{I}$ is incoherent.

  \item \begin{enumerate}
        \item Contractivity under all the incoherent completely trace preserving(ICTP) maps $\Phi_{ICTP}$, i.e., $C(\rho)\ge C(\Phi_{ICTP}(\rho))$ where $\Phi_{ICTP}(\rho)=\sum_n K_n\rho K_n^\dagger$ and $\{K_n\}$ is a set of Kraus operators, which satisfies $\sum_nK_n^\dagger K_n = \mathbb{I}$ with $K_n \mathcal{I} K_n^\dagger\subset \mathcal{I}$.

       \item Contractivity under selective measurement on average, i.e., $C(\rho)\ge \sum_n p_n C(\rho_n)$ where $\rho_n= K_n\rho K_n^\dagger/p_n$ and $p_n= Tr(K_n\rho K_n^\dagger)$ for any $\{K_n\}$ such that  $\sum_nK_n^\dagger K_n = \mathbb{I}$ and $K_n\mathcal{I} K_n^\dagger\subset \mathcal{I}, \quad \forall n$.
       \end{enumerate}

  \item Convexity, i.e., non-increasing  under mixing of quantum states : $\sum_n p_n C(\rho_n)\ge C(\sum_n p_n \rho_n)$  for any ensemble $\{p_n,\rho_n\}$.

\end{enumerate}
There are various ways to quantify coherence \cite{coh1,coh2,coh3,coh4}.
Here, we will mainly use a distance based measures of coherence:- $l_1$-norm of coherence
. $l_1$-norm coherence is defined in an intuitive way, via the off diagonal elements
of a density matrix $\rho$ in the reference basis,
\begin{equation}
C_{l_1}(\rho)=\sum_{i\ne j}|\rho_{ij}|.
\end{equation}
$l_1$-norm coherence satisfies all the properties of coherence measure.

\subsection{Residual coherence}
In this subsection, we analyze the relationship between coherence of global system and its subsystems of a bipartite state and define a new quantity namely ``residual coherence".  It quantifies the coherence which is inherent to the system and can not be shared locally. Let us consider a general two qubit state,
\begin{equation}
\rho_{ab}=\frac{1}{4}(I\otimes I +\sum_{i=1}^{3}x_i \sigma_i\otimes I +\sum_{i=1}^{3}y_i I\otimes \sigma_i+\sum_{i,j=1}^{3}t_{ij} \sigma_i\otimes \sigma_j).
\end{equation}
Now, the local and global $l_1$-coherence of this general state can be obtained as,
\begin{equation*}
\begin{split}
&C_{l_1}(\rho_{ab})=\frac{1}{2}(\sqrt{(y_1^2+t_{31}^2)+(y_2^2+t_{32}^2)}+\\
&\sqrt{(y_1^2-t_{31}^2)+(y_2^2-t_{32}^2)}+\sqrt{(x_1^2+t_{13}^2)+(x_2^2+t_{23}^2)}+\\
&\sqrt{(x_1^2-t_{13}^2)+(x_2^2-t_{23}^2)}+\sqrt{(t_{11}^2+t_{22}^2)+(t_{12}^2-t_{21}^2)}+\\
&\sqrt{(t_{11}^2-t_{22}^2)+(t_{12}^2+t_{21}^2)}),\\
\end{split}
\end{equation*}
\begin{equation*}
\begin{split}
C_{l_1}(\rho_{a})=\sqrt{x_1^2+x_2^2},\\
C_{l_1}(\rho_{b})=\sqrt{y_1^2+y_2^2}.\\
\end{split}
\end{equation*}
Some simple calculation reveals that for such arbitrary two qubit state,
$C_{l_1}(\rho_{ab})\ge C_{l_1}(\rho_{a})+C_{l_1}(\rho_{b})$.
Thus, the quantity $C_{l_1}(\rho_{ab})- C_{l_1}(\rho_{a})-C_{l_1}(\rho_{b})$ serves as a bona fide measure of `residual $l_1$-coherence' or simply `residual coherence' and we will denote it by $\delta_{l_1}(\rho_{ab})$.
%\begin{equation}\label{coherence score}
%\delta_{l_1}(\rho_{ab})=C_{l_1}(\rho_{ab})- C_{l_1}(\rho_{a})-C_{l_1}(\rho_{b}).
%\end{equation}

\section{Quantum cloning and deletion}\label{clone-del}
Before going to provide role of two seemingly different quantum processes,
quantum cloning and deletion, with the consumption and generation of quantum coherence,
we will first briefly describe different types quantum cloning and deletion machines.
\subsection{Universal quantum cloner}
We start with general form of cloning transformation $U$ on the Hilbert space $\mathcal{H}^2\otimes\mathcal{H}^2\otimes\mathcal{H}^x$, where $x$ is the dimension of the Hilbert space for ancilla states \cite{optimal cloner}:
\begin{eqnarray}\label{general cloner}
\begin{split}
U|0\rangle_{a}|0\rangle_{b}|X\rangle_x&=\hat{a}|00\rangle_{ab}|A\rangle_x+b_1|01\rangle_{ab}|B_1\rangle_x\\
&+b_2|10\rangle_{ab}|B_2\rangle_x+c|11\rangle_{ab}|C\rangle_x,\\
U|1\rangle_{a}|0\rangle_{b}|X\rangle_x&=\tilde{\hat{a}}|00\rangle_{ab}|\tilde{A}\rangle_x+\tilde{b_1}|01\rangle_{ab}|\tilde{B_1}\rangle_x\\
&+\tilde{b_2}|10\rangle_{ab}|\tilde{B_2}\rangle_x+\tilde{c}|11\rangle_{ab}|\tilde{C}\rangle_x
\end{split}
\end{eqnarray}
$|X\rangle$ is the initial state of ancilla and $A,B_i,C...$ refers to output ancilla states. Orthogonality condition is not imposed on $|A\rangle, |B_i\rangle$ and only they have to satisfy normality condition. We consider the coefficients $\hat{a}, b_i, c,...$ are in general complex and consider their free phases as $\hat{a}=|\hat{a}|e^{i\delta_{\hat{a}}},b_i=|b_i|e^{i\delta_{b_i}}, c=|c|e^{i\delta_{c}} $  and similar for other coefficients such that $|\hat{a}|=|\tilde{\hat{a}}|, |b_i|=|\tilde{b_i}|, |c|=|\tilde{c}|$. These coefficients satisfy the normalization conditions $|\hat{a}|^2+|b_1|^2+|b_2|^2+|c|^2=1$ and $|\tilde{\hat{a}}|^2+|\tilde{b_1}|^2+|\tilde{b_2}|^2+|\tilde{c}|^2=1$. If we want to clone an unknown state $|\psi\rangle=\alpha|0\rangle+\beta|1\rangle$,
where $ \alpha,\beta$ satisfy the relation $|\alpha|^2+|\beta|^2=1$, an isotropic universal symmetric quantum cloner must fulfill the conditions
(1) $\rho_a=\rho_b$ (symmetry), (2a) $\vec{s}_a=\eta_{\psi}\vec{s}_{\psi}$ (orientation invariance of Bloch vector) and (2b) Fidelity $F=\text{Tr}(\rho_{\psi}\rho_1)=\frac{1}{2}(1+\eta_{\psi})=\text{constant}$, where $\rho_a,\rho_b$ are the reduced local output states and $\vec{s}_a,\vec{s}_{\psi}$ are Bloch vectors corresponding to the states $\rho_a, |\psi\rangle\langle\psi|$.
The explicit form of reduction factor $\eta$  and fidelity $F$ are
$$\eta=\eta_{\psi}=|\hat{a}|^2-|c|^2 , F=\frac{1}{2}(1+\eta).$$
In our work, we will consider in particular two state independent isotropic and symmetric quantum cloners: optimal universal quantum cloner and phase covariant quantum cloner.\\

\paragraph{\textbf{Optimal universal quantum cloner:}}
The optimal universal quantum cloning machine(OUQC) was introduced by Bruss et.al. \cite{optimal cloner}. The action of this machine is given by the following transformation
\begin{widetext}
\begin{equation}\label{optimal cloner}
\begin{split}
&U|0\rangle_a|0\rangle_b|X\rangle_x=\sqrt{\frac{2}{3}}e^{i\delta_{\hat{a}}}|00\rangle_{ab}|A\rangle_x+\sqrt{\frac{1}{6}}e^{i\delta_{\tilde{\hat{a}}}}(|01\rangle_{ab}+|10\rangle_{ab})|A_{\perp}\rangle_{x},\\
&U|1\rangle_{a}|0\rangle_{b}|X\rangle_{x}=\sqrt{\frac{2}{3}}e^{i\delta_{\tilde{\hat{a}}}}|11\rangle_{ab}|A_{\perp}\rangle_x+\sqrt{\frac{1}{6}}e^{i\delta_{\hat{a}}}(|01\rangle_{ab}+|10\rangle_{ab})|A\rangle_{x},
\end{split}
\end{equation}
\end{widetext}
with $\langle A|A_{\perp}\rangle=0$. $a,b$ and $x$ denote qubits corresponding to the input-state, blank state and machine state respectively. This is an input state independent $1\rightarrow 2$ quantum cloner with optimal cloning fidelity $\frac{5}{6}$.\\

\paragraph{\textbf{Phase-covariant cloner:}}
The phase covariant cloner was introduced by Bruss et al. in \cite{Phase-cov cloner}. The action of this cloner is described  by the following operation
\begin{widetext}
\begin{equation}\label{phase-cov cloner}
\begin{split}
&U|0\rangle_{a}|0\rangle_{b}|X\rangle_x=(\frac{1}{2}+\sqrt{\frac{1}{8}})|00\rangle_{ab}|0\rangle_x+\sqrt{\frac{1}{8}}(|01\rangle_{ab}+|10\rangle_{ab})|1\rangle_x+(\frac{1}{2}-\sqrt{\frac{1}{8}})|11\rangle_{ab}|0\rangle_x,\\
&U|1\rangle_{a}|0\rangle_{b}|X\rangle_x=(\frac{1}{2}+\sqrt{\frac{1}{8}})|11\rangle_{ab}|1\rangle_x+\sqrt{\frac{1}{8}}(|10\rangle_{ab}+|01\rangle_{ab})|0\rangle_x+(\frac{1}{2}-\sqrt{\frac{1}{8}})|00\rangle_{ab}|1\rangle_x.
\end{split}
\end{equation}
\end{widetext}
This transformations are optimal for a restricted class of input states of the form
$|\psi\rangle_{\phi}=\frac{1}{\sqrt{2}}(|0\rangle+e^{i \phi}|1\rangle)$, where $\phi\in[0,2\pi]$. But in context of physical implementation, this restriction of input states is well motivated because all existing quantum cryptography experiments were done using the states that are on the equator, rather than the states that span the whole Bloch sphere.\\

\subsection{Deleting Machine}
The complementary to the `quantum no-cloning theorem' is the `quantum no-deleting'
principle \cite{deleting machine}. It states that linearity of quantum theory forbids deletion of
one unknown quantum state against a copy in either a reversible or an irreversible
manner. Here we consider the PB-deleting machine defined by Pati et al. in \cite{deleting machine}. Let us consider a qubit pure state $|\psi\rangle$ as our initial state.
\begin{equation}\label{pure state}
|\psi\rangle=\alpha|0\rangle+\beta|1\rangle,
\end{equation}
where $ \alpha,\beta$ satisfy $|\alpha|^2+|\beta|^2=1.$ The $N$ copies of qubit (\ref{pure state}) can be written as
$$|\psi\rangle^{\otimes N}=\alpha^N|0\rangle^{\otimes N}+\beta^N|1\rangle^{\otimes N}+\sum_{k=1}^{N-1}f_k(\alpha,\beta)|k\rangle,$$
where $|k\rangle$'s are $(N-1)$ orthogonal bit string states living in symmetric subspace.

This $N-\text{to}-M$ quantum deleting machine for orthogonal qubits is defined by
\begin{equation}
\begin{split}
&|0\rangle^{\otimes N}|X\rangle\rightarrow |0\rangle^{\otimes M}|\Sigma\rangle^{\otimes(N-M)}|X_0\rangle,\\
&|1\rangle^{\otimes N}|X\rangle\rightarrow |1\rangle^{\otimes M}|\Sigma\rangle^{\otimes(N-M)}|X_1\rangle,\\
&|k\rangle|X\rangle\rightarrow |k'\rangle
\end{split}
\end{equation}
where $|\Sigma\rangle$ is the blank state and $|k'\rangle$ is the final state of symmetric N qubits and ancilla.

\section{Cloning followed by deletion}
\begin{figure}[htb]
\subfloat[]{\includegraphics[scale=0.5]{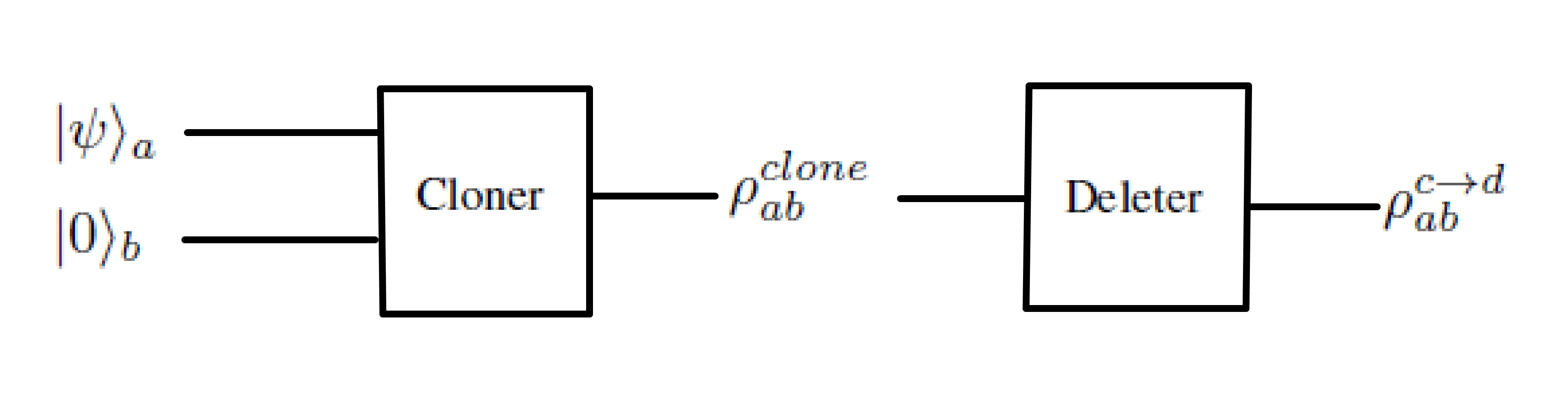}\label{ctod}}\\
\subfloat[]{\includegraphics[scale=0.4]{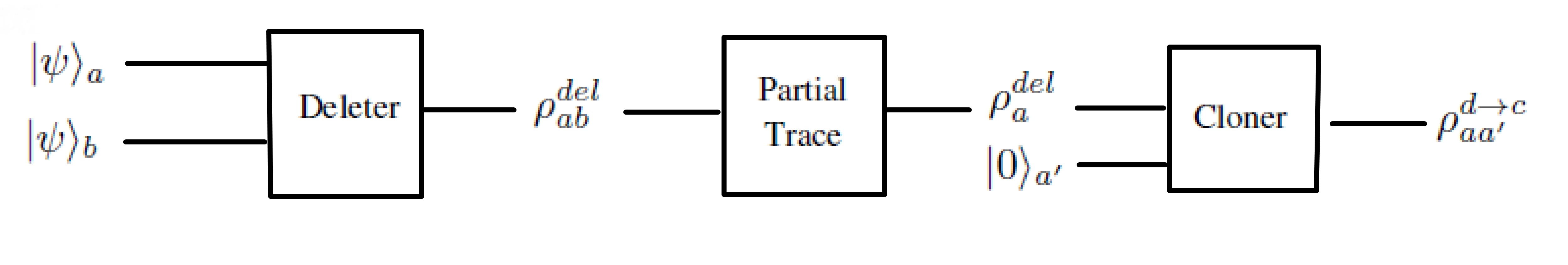}\label{dtoc}}
\caption{Fig(\ref{ctod}) describes the cloning process followed by deletion and Fig(\ref{dtoc}) describes the deletion process followed by cloning.}
\label{quantum process}
\end{figure}

We consider a quantum process (ref. FIG (\ref{ctod})) in which a system undergoes cloning operation and then deletion operation on the output system. The input state $\rho_{ab}^{in}=|\psi\rangle_a\langle\psi|\otimes|0\rangle_b\langle 0|$ is acted on by the cloning machine (\ref{general cloner}). The output cloned state is given by (ref. Appendix  \ref{expression})
\begin{equation}\label{general cloned states}
\rho^{clone}_{ab}=\sum_{i,j,k,l=0}^1p_{ij,kl}|ij\rangle\langle kl|.
\end{equation}
The coherence of the input state and the cloned states, corresponding to the global and local systems are  of the form
\begin{equation}\label{initial coh}
\begin{split}
&C_{l_1}(\rho_{ab}^{in})=2|\alpha \beta|,\\
&C_{l_1}(\rho_{a}^{in})=2|\alpha \beta|,\\
&C_{l_1}(\rho_{b}^{in})=0
\end{split}
\end{equation}
The initial residual coherence is zero. But after cloning operation the local and global coherence become,
\begin{eqnarray}
\begin{split}\label{cloned coh}
C_{l_1}(\rho_{ab}^{clone})&=2(|p_{00,01}|+|p_{00,10}|+|p_{00,11}|+\\&|p_{01,10}|+|p_{01,11}|+|p_{10,11}|)\\
C_{l_1}(\rho_{a}^{clone})&=2(|p_{00,10}+p_{01,11}|)=\eta|\alpha\beta|,\\
C_{l_1}(\rho_{b}^{clone})&=2(|p_{00,01}+p_{10,11}|)=\eta|\alpha\beta|.
\end{split}
\end{eqnarray}
Corresponding to the optimal cloner, coherence of the global and local systems can be obtained as,
\begin{equation}\label{oc}
\begin{split}
C_{l_1}^{oc}(\rho_{ab}^{clone})&=\frac{8}{3}|\alpha \beta|+\frac{1}{3},\\
C_{l_1}^{oc}(\rho_{a}^{clone})&=C_{l_1}(\rho_{b}^{clone})=\frac{4}{3}|\alpha \beta| ,\\
\end{split}
\end{equation}
Hence residual coherence $\delta_{l_1}^{oc}(\rho_{ab}^{clone})=\frac{1}{3}$
and the similar quantities for phase covariant cloning operation are
\begin{equation}\label{pc}
\begin{split}
C_{l_1}^{pc}(\rho_{ab}^{clone})&=2\sqrt{2}|\alpha \beta|+\frac{1}{2},\\
C_{l_1}^{pc}(\rho_{a}^{clone})&=C_{l_1}(\rho_{b}^{clone})=\sqrt{2}|\alpha \beta|.\\
\end{split}
\end{equation}
Hence residual coherence $\delta_{l_1}^{pc}(\rho_{ab}^{clone})=\frac{1}{2}.$
Therefore, after cloning operation residual coherence increases and freezes to a certain level independent of the initial state. However, phase covariant cloner generates much residual coherence than optimal cloner.  From the expression of residual coherence, it is evident that due to consideration of symmetric cloner the residual coherence is independent of the input states.

Now a state dependent deleting operation is applied on imperfectly cloned copies (\ref{general cloned states}). The action of the deleting machine is given by the following unitary operations (\cite{deleting machine}),
\begin{equation}\label{deletion general imperfect}
\begin{split}
& |00\rangle_{ab}|A\rangle_x \rightarrow |00\rangle_{ab}|A_0\rangle_x\\
& |00\rangle_{ab}|\tilde{C}\rangle_x \rightarrow |00\rangle_{ab}|A_1\rangle_x\\
& |11\rangle_{ab}|\tilde{A}\rangle_x \rightarrow |10\rangle_{ab}|A_2\rangle_x\\
& |11\rangle_{ab}|C\rangle_x \rightarrow |10\rangle_{ab}|A_3\rangle_x\\
& |01\rangle_{ab}|B_1\rangle_x \rightarrow |01\rangle_{ab}|B_1\rangle_x\\
& |10\rangle_{ab}|B_2\rangle_x \rightarrow |10\rangle_{ab}|B_2\rangle_x\\
& |01\rangle_{ab}|\tilde{B_2}\rangle_x \rightarrow |01\rangle_{ab}|\tilde{B_2}\rangle_x\\
& |10\rangle_{ab}|\tilde{B_1}\rangle_x \rightarrow |10\rangle_{ab}|\tilde{B_1}\rangle_x\\
\end{split}
\end{equation}
The imperfectly deleted state is given by, (refer Appendix \ref{expression})
\begin{equation}\label{general cloned del state}
\begin{split}
&\rho^{c\rightarrow d}_{ab}=r_{00,00}|00\rangle\langle 00|+r_{01,01}|01\rangle\langle01|+r_{10,10}|10\rangle\langle10|+\\
&r_{00,01}|00\rangle\langle01|+r_{00,10}|00\rangle\langle10|+r_{01,00}|01\rangle\langle00|+\\
&r_{01,10}|01\rangle\langle10|+r_{10,00}|10\rangle\langle00|+r_{10,01}|10\rangle\langle01|.
\end{split}
\end{equation}
Coherence of this output state is given by,
\begin{equation}
\begin{split}
C_{l_1}(\rho_{ab}^{c\rightarrow d})&=2(|r_{00,01}|+|r_{00,10}|+|r_{10,01}|),\\
C_{l_1}(\rho_{a}^{c\rightarrow d})&=2|r_{00,10}|,\\
 C_{l_1}(\rho_{b}^{c\rightarrow d})&=2|r_{00,01}|.
\end{split}
\end{equation}
It reveals that the residual coherence of the output deleted system is non-zero. Corresponding to the optimal cloner and phase covariant cloner the above quantities can be explicitly written as,
\begin{equation}
\begin{split}
C_{l_1}^{oc}(\rho_{ab}^{c\rightarrow d})&=\frac{1}{3},\\
C_{l_1}^{oc}(\rho_a^{c\rightarrow d})&=C_{l_1}(\rho_b^{c\rightarrow d})=0,\\
\Rightarrow\delta_{l_1}^{oc} (\rho_{ab}^{c\rightarrow d})&=\frac{1}{3},
\end{split}
\end{equation}
\begin{equation}
\begin{split}
C_{l_1}^{pc}(\rho_{ab}^{c\rightarrow d})&=\frac{1}{4},\\
C_{l_1}^{pc}(\rho_a^{c\rightarrow d})&=C_{l_1}(\rho_b^{c\rightarrow d})=0,\\
\Rightarrow\delta_{l_1}^{pc}(\rho_{ab}^{c\rightarrow d})&=\frac{1}{4}.
\end{split}
\end{equation}
It is evident that the cloning operation increases coherence of the blank part,( i.e., the local system of b) and global system $\rho_{ab}$ but decreases coherence of the system to be copied, i.e., local system of a. However, the global increment is greater than the total local increment (i.e., increment in subsystem b - decay in subsystem a), i.e., cloning is a coherence generating process. The generated coherence in the cloning process is greater in phase covariant cloner than optimal cloner. On the other hand, deletion decreases the coherence in both the local subsystems and global system. Thus, deletion is a decohering  process. After deletion, relation between global decay and total local decay depends on the initial cloning machine. In optimal cloner, global decay is same as total local decay but in phase covariant cloner global decay is greater than total local decay. The fact indicates that global coherence is not merely distributed among subsystems. This phenomenon occurs due to the presence of more information in phase covariant cloning transformation than optimal cloning transformation. We also observe that the cloning operation increases the residual coherence but under the deletion operation it either remains constant or is decreased.

The amount of consumption or generation of coherence in this whole process is given by the difference between the amount of coherence of final output state and amount of coherence of initial input state. We denote this difference as $\Delta C^{c\rightarrow d}=C_{l_1}(\rho_{ab}^{c\rightarrow d})-C_{l_1}(\rho_{ab}^{in})$. We also denote the difference between the amount of residual coherence of final output and initial input states as $\Delta \delta^{c\rightarrow d}=\delta_{l_1}(\rho_{ab}^{c\rightarrow d})-\delta_{l_1}(\rho_{ab}^{in})$.  These two quantities for optimal cloner and phase covariant cloner are respectively
\begin{equation}
\begin{split}
\Delta C^{c\rightarrow d}_{oc}=\frac{1}{3}-2|\alpha \beta|,\\
\Delta \delta^{c\rightarrow d}_{oc}=\frac{1}{3},\\
\Delta C^{c\rightarrow d}_{pc}=\frac{1}{4}-2|\alpha \beta|,\\
\Delta \delta^{c\rightarrow d}_{oc}=\frac{1}{4}.\\
\end{split}
\end{equation}
\begin{figure}[htb]
\includegraphics[scale=0.65]{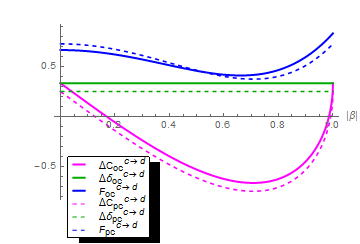}
\caption{Fidelity(blue), residual coherence difference(green),
coherence difference(magenta) w.r.t state parameter $|\beta|$ for optimal
cloner (solid lines) and phase covariant cloner(dashed lines) in cloning followed
by deletion process. Coherence consumption and generation during the process are
evident from the figure. While the residual coherence remains fixed throughout
the process, fidelity and coherence difference show similar monotonic behavior.  }
\label{ctd1}
\end{figure}
In the whole process, for optimal cloner, coherence is consumed  when $0.169102\le |\beta|\le 0.985598$ and for phase covariant cloner, coherence
consumption occurs for the states with $0.126004\le |\beta|\le 0.99203$. For rest of the range of $|\beta|$ coherence generation occurs  (refer Fig. \ref{ctd1}). The fidelity of this process is the overlap of the final output state $\rho^{c\rightarrow d}_{ab}$ with the initial state $\rho^{in}_{ab}$ and for optimal cloner the fidelity is given by $$F^{c\rightarrow d}_{oc}=\frac{2}{3}(1-2|\alpha \beta|^2)+\frac{1}{6}|\beta|^2$$
whereas for phase covariant cloner, it is given by $$F^{c\rightarrow d}_{pc}=\frac{1}{8\sqrt{2}}(4+3\sqrt{2}-16|\alpha \beta|^2).$$ We have the following the bounds : $\frac{79}{192}\le F^{c\rightarrow d}_{oc} \le \frac{5}{6}$ and $\frac{3}{8}\le F^{c\rightarrow d}_{pc} \le \frac{4+3\sqrt{2}}{8\sqrt{2}}$.

\section{Deletion followed by cloning}\label{del-clone}
Next, we consider the scenario (ref. FIG (\ref{dtoc})) where we will first perform deletion operation and then the cloning operation on the outcome state. This process is just the reversal of the previous procedure (\ref{clone-del}). We take two copies of the pure state $|\psi\rangle$  as  the initial state of this process.
\begin{equation}\label{sigma initial state}
  \rho_{ab}^{0}=|\psi\rangle_a\langle\psi|\otimes|\psi\rangle_b\langle\psi|,
  \end{equation}
and  the following state dependent deleting operation\cite{deleting machine} is applied on it,
\begin{equation}\label{deleting machine}
   \begin{split}
   & |00\rangle_{ab}|A\rangle_x \rightarrow |00\rangle_{ab}|Q_0\rangle_x,\\
   & |11\rangle_{ab}|A\rangle_x \rightarrow |10\rangle_{ab}|Q_1\rangle_x,\\
   & (|01\rangle_{ab}+|10\rangle_{ab})|A\rangle_x \rightarrow (|01\rangle_{ab}+|10\rangle_{ab})|A\rangle_x,\\
   \end{split}
   \end{equation}
where $\langle Q_i|Q_j\rangle = 0$. Applying deleting machine (\ref{deleting machine}) on $\rho_{ab}^{0}$, we get the imperfectly deleted copies,
\begin{equation}\label{imperfect deleted}
  \rho_{ab}^{ del}=\alpha^4|00\rangle_{ab}\langle 00|+\beta^4|10\rangle_{ab}\langle 10|+2\alpha^2\beta^2|\psi^+\rangle_{ab}\langle \psi^+|.
\end{equation}
The fidelity of deleted copies is given by $F_1=1-|\alpha\beta|^2$. Global and local coherence of the initial and the deleted copies can be obtained as,
\begin{equation}\label{del coh}
\begin{split}
C_{l_1}(\rho_{ab}^{0})&=4|\alpha \beta|(1+|\alpha \beta|),\\
C_{l_1}(\rho_a^{0})&=C_{l_1}(\rho_b^{0})=2|\alpha \beta|,\\
\Rightarrow\delta_{l_1}(\rho_{ab}^{0})&=4|\alpha \beta|^2\\
C_{l_1}(\rho_{ab}^{del})&=2|\alpha \beta|^2,\\
C_{l_1}(\rho_a^{del})&=C_{l_1}(\rho_b^{del})=0,\\
\Rightarrow\delta_{l_1}(\rho_{ab}^{del})&=2|\alpha \beta|^2.
\end{split}
\end{equation}
Hence the deletion operation decreases the coherence of the global as well as local systems.
 Next, the system undergoes general quantum cloning operation.
The cloning operation can be applied both on the reduced states $\rho_{a}^{del}$ and
$\rho_{b}^{del}$. We consider $|0\rangle_{a'}$ and $|0\rangle_{b'}$ as the blank states
for subsystems $a$ and $b$ respectively. Corresponding to the reduced states $\rho_{a}^{del}$ and $\rho_{b}^{del}$, we get two cloned output states,(ref. Appendix \ref{expression})
\begin{equation}\label{del-clone state}
\begin{split}
& \rho^{d\rightarrow c}_{aa'}=\sum_{i,j,k,l=0}^1m_{ij,kl}|ij\rangle\langle kl|,\\
&\rho^{d\rightarrow c}_{bb'}=\sum_{i,j,k,l=0}^1 n_{ij,kl}|ij\rangle\langle kl|.\\
\end{split}
\end{equation}
Coherence of the final states corresponding to two different initial states can be written as
\begin{equation}\label{del-clone coh for gen}
\begin{split}
C_{l_1}(\rho_{aa'}^{d\rightarrow c})= 2(|m_{00,01}|+|m_{00,10}|+&|m_{00,11}|+|m_{01,10}|+\\&|m_{01,11}|+|m_{10,11}|),\\
C_{l_1}(\rho_{a}^{d\rightarrow c})=C_{l_1}(\rho_{a'}^{d\rightarrow c})=0;\\
\end{split}
\end{equation}
and\\
\begin{equation}
\begin{split}
C_{l_1}(\rho_{bb'}^{d\rightarrow c})= 2(|n_{00,01}|+|n_{00,10}|+&|n_{00,11}|+|n_{01,10}|+\\&|n_{01,11}|+|n_{10,11}|),\\
C_{l_1}(\rho_{b}^{d\rightarrow c})=C_{l_1}(\rho_{b'}^{d\rightarrow c})=0.\\
\end{split}
\end{equation}
Above quantities for the optimal cloner and phase covariant cloner are respectively
\begin{equation}\label{del-clone coh}
\begin{split}
& C_{l_1}^{oc}(\rho_{a\acute{a}}^{d\rightarrow c})=C_{l_1}(\rho_{b\acute{b}}^{d\rightarrow c})=\frac{1}{3},\\
&C_{l_1}^{oc}(\rho_a^{d\rightarrow c})=C_{l_1}(\rho_b^{d\rightarrow c})=0,\\
\Rightarrow&\delta_{l_1}^{oc}(\rho_{a\acute{a}}^{d\rightarrow c})=\frac{1}{3};\\
& C_{l_1}^{pc}(\rho_{a\acute{a}}^{d\rightarrow c})=C_{l_1}(\rho_{b\acute{b}}^{d\rightarrow c})=\frac{1}{2},\\
&C_{l_1}^{pc}(\rho_a^{d\rightarrow c})=C_{l_1}(\rho_b^{d\rightarrow c})=0,\\
\Rightarrow&\delta_{l_1}^{pc}(\rho_{a\acute{a}}^{d\rightarrow c})=\frac{1}{2}.
\end{split}
\end{equation}
Initially, after deletion, coherence of both global and local systems decreases and the global decay is greater than total local decay. This deleting machine produces zero coherence in subsystems but positive coherence in global system. Cloning can not change the coherence in subsystems but freezes the global coherence. The coherence after optimal cloning operation is not greater than the coherence of deleted copy in whole range of state parameter but for phase covariant cloner it is greater in the whole range. The residual coherence in phase covariant cloner is greater than that in optimal cloner.

The fidelity of this process for optimal and phase covariant cloner are
given by as follows $$F_{oc}^{d\rightarrow c}=\frac{2}{3}(1-2|\alpha \beta|^2),$$ and
$$F_{pc}^{d\rightarrow c}=\frac{1}{8\sqrt{2}}(3\sqrt{2}+4-(16-2\sqrt{2})|\alpha \beta|^2).$$
Similar to the first process, the difference between the amount of coherence of the final
output state $\rho_{aa'}^{d\rightarrow c}$ and the amount of coherence of the initial
input state $\rho_{ab}^{0}$ has been denoted as $\Delta C^{d\rightarrow c}$.
Similarly the difference between the amount of residual coherence of the final and
initial states has been denoted by $\Delta\delta^{d\rightarrow c}$.
The explicit expressions of these two types of differences for optimal cloner and phase covariant cloner are given by
\begin{equation}
 \begin{split}
  \Delta C_{oc}^{d\rightarrow c}=\frac{1}{3}-4|\alpha\beta|(1+|\alpha\beta|),\\
  \Delta \delta_{oc}^{d\rightarrow c}=\frac{1}{3}-4|\alpha\beta|^2,\\
  \Delta C_{pc}^{d\rightarrow c}=\frac{1}{2}-4|\alpha\beta|(1+|\alpha\beta|),\\
  \Delta \delta_{pc}^{d\rightarrow c}=\frac{1}{2}-4|\alpha\beta|^2.
 \end{split}
\end{equation}
\begin{figure}[htb]
 \includegraphics[scale=0.65]{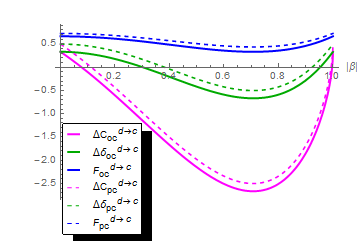} 
 \caption{Fidelity(blue), residual coherence difference(green), coherence difference(magenta) w.r.t state parameter
$|\beta|$ for optimal cloner (solid lines) and phase covariant cloner(dashed lines)
in deletion followed by cloning process. Coherence consumption and generation are evident
in the process. Both the residual coherence and fidelity have similar proportional behavior to coherence difference. Coherence consumption decreases the fidelity whereas the generation indicates fidelity enhancement. }
\label{dtc1}
\end{figure}
In the whole process, for optimal cloner, coherence is consumed  for the states with
$0.077758\le |\beta|\le 0.996986$ and for phase covariant cloner, coherence consumption
occurs for the states with $0.113098\le |\beta|\le 0.993584$ (refer Fig. \ref{dtc1}).
Coherence generation occurs in rest of the range of $|\beta|$.

\section{Conclusion}
In this paper we have analyzed $l_1$-norm coherence under two quantum processes- cloning followed by deletion and deletion followed by cloning. We have observed that under cloning, coherence of cloned subsystem decreases but coherence of
blank subsystem and global system increase. Again global increment of coherence is
greater than total local increment, i.e., cloning is basically a coherence generating process.
On the other hand, deletion decreases coherence in both subsystems and global system,
i.e., deletion is decohering process. On the other side, cloning increases the residual
coherence but deletion decreases it. Thus the performance of cloning and deletion with
coherence consumption and generation is established. Even the fidelities of the processes
 have similar monotonic behavior with the generation or consumption of
coherence through each process.
We hope our results will enable
to find new insight to understand quantum coherence as an important resource.

\begin{acknowledgements}
	The author S. Karmakar acknowledges the financial support from UGC, India. The author A.Sen acknowledges  NBHM, DAE India and the author D. Sarkar  acknowledges SERB, DST  India and DSA-SAP for financial support.
\end{acknowledgements}

\begin{appendix}
\begin{widetext}
\section{Explicit expressions}\label{expression}
Explicit expression of the  coefficients of the  cloned state $\rho_{ab}^{clone}$ given in Eq.(\ref{general cloned states}):
\begin{equation}\label{expression for pi}
\begin{split}
& p_{00,00}=\alpha^2\hat{a}^2+\beta^2\tilde{c}^2+\alpha\beta(\hat{a}\tilde{c}^*\langle\tilde{C}|A\rangle+\hat{a}^*\tilde{c}\langle A|\tilde{C}\rangle)=\langle u_1|u_1\rangle\\
& p_{01,01}=\alpha^2b_1^2+\beta^2\tilde{b_2}^2+\alpha\beta(b_1\tilde{b_2}^*\langle\tilde{B_2}|B_1\rangle+b_1^*\tilde{b_2}\langle B_1|\tilde{B_2}\rangle)=\langle v_1|v_1\rangle\\
& p_{10,10}=\alpha^2b_2^2+\beta^2\tilde{b_1}^2+\alpha\beta(b_2\tilde{b_1}^*\langle\tilde{B_1}|B_2\rangle+b_2^*\tilde{b_1}\langle B_2|\tilde{B_1}\rangle)=\langle v_2|v_2\rangle\\
& p_{11,11}=\alpha^2c^2+\beta^2\tilde{\hat{a}}^2+\alpha\beta(c\tilde{\hat{a}}^*\langle\tilde{A}|C\rangle+c^*\tilde{\hat{a}}\langle C|\tilde{A}\rangle)=\langle u_2|u_2\rangle\\
& p_{00,01}=p^*_{01,00}=\alpha^2\hat{a}b_1^*\langle B_1|A\rangle+\beta^2\tilde{b_2}^*\tilde{c}\langle \tilde{B_2}|\tilde{C}\rangle+\alpha\beta(\hat{a}\tilde{b_2}^*\langle\tilde{B_2}|A\rangle+b_1^*\tilde{c}\langle B_1|\tilde{C}\rangle)=\langle v_1|u_1\rangle\\
& p_{00,10}=p^*_{10,00}=\alpha^2\hat{a}b_2^*\langle B_2|A\rangle+\beta^2\tilde{b_1}^*\tilde{c}\langle \tilde{B_1}|\tilde{C}\rangle+\alpha\beta(\hat{a}\tilde{b_1}^*\langle\tilde{B_1}|A\rangle+b_2^*\tilde{c}\langle B_2|\tilde{C}\rangle)=\langle v_2|u_1\rangle\\
& p_{00,11}=p^*_{11,00}=\alpha^2\hat{a}c^*\langle C|A\rangle+\beta^2\tilde{\hat{a}}^*\tilde{c}\langle \tilde{A}|\tilde{C}\rangle+\alpha\beta(\hat{a}\tilde{\hat{a}}^*\langle\tilde{A}|A\rangle+c^*\tilde{c}\langle C|\tilde{C}\rangle)=\langle u_2|u_1\rangle\\
& p_{01,10}=p^*_{10,01}=\alpha^2b_1b_2^*\langle B_2|B_1\rangle+\beta^2\tilde{b_1}^*\tilde{b_2}\langle \tilde{B_1}|\tilde{B_2}\rangle+\alpha\beta(b_1\tilde{b_1}^*\langle\tilde{B_1}|B_1\rangle+b_2^*\tilde{b_2}\langle B_2|\tilde{B_2}\rangle)=\langle v_2|v_1\rangle\\
& p_{01,11}=p^*_{11,01}=\alpha^2b_1c^*\langle C|B_1\rangle+\beta^2\tilde{\hat{a}}^*\tilde{b_2}\langle \tilde{A}|\tilde{B_2}\rangle+\alpha\beta(b_1\tilde{\hat{a}}^*\langle\tilde{A}|B_1\rangle+c^*\tilde{b_2}\langle C|\tilde{B_2}\rangle)=\langle u_2|v_1\rangle\\
& p_{10,11}=p^*_{11,10}=\alpha^2b_2c^*\langle C|B_2\rangle+\beta^2\tilde{\hat{a}}^*\tilde{b_1}\langle \tilde{A}|\tilde{B_1}\rangle+\alpha\beta(b_2\tilde{\hat{a}}^*\langle\tilde{A}|B_2\rangle+c^*\tilde{b_1}\langle C|\tilde{B_1}\rangle)=\langle u_2|v_2\rangle\\
\end{split}
\end{equation}
where  $|u_1\rangle=\alpha \hat{a}|A\rangle+\beta \tilde{c}|\tilde{C}\rangle$, $|u_2\rangle=\alpha c|C\rangle+\beta \tilde{\hat{a}}|\tilde{A}\rangle$, $|v_1\rangle=\alpha b_1|B_1\rangle+\beta \tilde{b_2}|\tilde{B_2}\rangle$, $|v_2\rangle=\alpha b_2|B_2\rangle+\beta \tilde{b_1}|\tilde{B_1}\rangle$\\\\

Explicit expression of the  coefficients of the  state $\rho_{ab}^{c\rightarrow d}$ given in Eq.(\ref{general cloned del state}):
\begin{equation}\label{expression for ri}
\begin{split}
& r_{00,00}=\alpha^2\hat{a}^2+\beta^2\tilde{c}^2+\alpha\beta(\hat{a}\tilde{c}^*\langle A_1|A_0\rangle+\hat{a}^*\tilde{c}\langle A_0|A_1\rangle)\\
& r_{01,01}=\alpha^2b_1^2+\beta^2\tilde{b_2}^2+\alpha\beta(b_1\tilde{b_2}^*\langle\tilde{B_2}|B_1\rangle+b_1^*\tilde{b_2}\langle B_1|\tilde{B_2}\rangle)\\
& r_{10,10}=\alpha^2(b_2^2+c^2+b_2c^*\langle A_3|B_2\rangle+b_2^*c\langle B_2|A_3\rangle)+\beta^2(\tilde{\hat{a}}^2+\tilde{b_1}^2+\tilde{\hat{a}}^*\tilde{b_1}\langle A_2|\tilde{B_1}\rangle+\tilde{\hat{a}}\tilde{b_1}^*\langle \tilde{B_1}|A_2\rangle)\\
& +\alpha\beta(b_2\tilde{\hat{a}}^*\langle A_2|B_2\rangle+b_2^*\tilde{\hat{a}}\langle B_2|A_2\rangle+b_2\tilde{b_1}^*\langle\tilde{B_1}|B_2\rangle+b_2^*\tilde{b_1}\langle B_2|\tilde{B_1}\rangle+c^*\tilde{\hat{a}}\langle A_3|A_2\rangle+c\tilde{\hat{a}}^*\langle A_2|A_3\rangle+c^*\tilde{b_1}\langle A_3|\tilde{B_1}\rangle+c\tilde{b_1}^*\langle \tilde{B_1}|A_3\rangle)\\
& r_{00,01}=r^*_{01,00}=\alpha^2\hat{a}b_1^*\langle B_1|A_0\rangle+\beta^2\tilde{b_2}^*\tilde{c}\langle \tilde{B_2}|A_1\rangle+\alpha\beta(\hat{a}\tilde{b_2}^*\langle\tilde{B_2}|A_0\rangle+b_1^*\tilde{c}\langle B_1|A_1\rangle)\\
& r_{00,10}=r^*_{10,00}=\alpha^2(\hat{a}b_2^*\langle B_2|A_0\rangle+\hat{a}c^*\langle A_3|A_0\rangle)+\beta^2(\tilde{b_1}^*\tilde{c}\langle \tilde{B_1}|A_1\rangle+\tilde{\hat{a}}^*\tilde{c}\langle \tilde{A_2}|A_1\rangle)\\
& +\alpha\beta(\hat{a}\tilde{\hat{a}}^*\langle A_1|A_0\rangle+\hat{a}\tilde{b_1}^*\langle\tilde{B_1}|A_0\rangle+\hat{a}^*\tilde{c}\langle A_0|A_1\rangle+b_2^*\tilde{c}\langle B_2|A_1\rangle)\\
& r_{01,10}=r^*_{10,01}=\alpha^2(b_1b_2^*\langle B_2|B_1\rangle+b_1c^*\langle A_3|B_2\rangle)+\beta^2(\tilde{b_1}^*\tilde{b_2}\langle \tilde{B_1}|\tilde{B_2}\rangle+\tilde{\hat{a}}^*\tilde{b_2}\langle A_2|\tilde{B_2}\rangle)\\
& +\alpha\beta(b_1\tilde{\hat{a}}^*\langle A_2|B_1\rangle+b_1\tilde{b_1}^*\langle\tilde{B_1}|B_1\rangle+b_2^*\tilde{b_2}\langle B_2|\tilde{B_2}\rangle+c^*\tilde{b_2}\langle A_3|\tilde{B_2}\rangle)\\
\end{split}
\end{equation}\\\\

Explicit expression of the  coefficients of the state $\rho_{aa'}^{d\rightarrow c}$ given in Eq.(\ref{del-clone state}):
\begin{equation}\label{expression for mi}
\begin{split}
& m_{00,00}=\alpha^2\hat{a}^2+\beta^2\tilde{c}^2\\
& m_{01,01}=\alpha^2b_1^2+\beta^2\tilde{b_2}^2\\
& m_{10,10}=\alpha^2b_2^2+\beta^2\tilde{b_1}^2\\
& m_{11,11}=\alpha^2c^2+\beta^2\tilde{\hat{a}}^2\\
& m_{00,01}=m^*_{01,00}=\alpha^2\hat{a}b_1^*\langle B_1|A\rangle+\beta^2\tilde{b_2}^*\tilde{c}\langle \tilde{B_2}|\tilde{C}\rangle\\
& m_{00,10}=m^*_{10,00}=\alpha^2\hat{a}b_2^*\langle B_2|A\rangle+\beta^2\tilde{b_1}^*\tilde{c}\langle \tilde{B_1}|\tilde{C}\rangle\\
& m_{00,11}=m^*_{11,00}=\alpha^2\hat{a}c^*\langle C|A\rangle+\beta^2\tilde{\hat{a}}^*\tilde{c}\langle \tilde{A}|\tilde{C}\rangle\\
& m_{01,10}=m^*_{10,01}=\alpha^2b_1b_2^*\langle B_2|B_1\rangle+\beta^2\tilde{b_1}^*\tilde{b_2}\langle \tilde{B_1}|\tilde{B_2}\rangle\\
& m_{01,11}=m^*_{11,01}=\alpha^2b_1c^*\langle C|B_1\rangle+\beta^2\tilde{\hat{a}}^*\tilde{b_2}\langle \tilde{A}|\tilde{B_2}\rangle\\
& m_{10,11}=m^*_{11,10}=\alpha^2b_2c^*\langle C|B_2\rangle+\beta^2\tilde{\hat{a}}^*\tilde{b_1}\langle \tilde{A}|\tilde{B_1}\rangle\\
\end{split}
\end{equation}
Explicit expression of the  coefficients of the state $\rho_{bb'}^{d\rightarrow c}$ given in Eq.(\ref{del-clone state}) are same as $m_i$'s, just replacing $\alpha^2$ by $1-\alpha^2\beta^2$ and $\beta^2$ by $\alpha^2\beta^2$ in Eqs.(\ref{expression for mi}).

 \end{widetext}
\end{appendix}
\end{document}